\documentclass[sigconf,table,nonacm]{acmart}

\usepackage[utf8]{inputenc}
\usepackage{amsmath}
\usepackage{graphicx}
\usepackage{siunitx}
\usepackage{array}
\usepackage{listings}
\usepackage[frozencache]{minted}
\usepackage{soulutf8}
\usepackage{tikz}
\usetikzlibrary{decorations.pathreplacing}
\usepackage{float}
\usepackage{tabularx}
\newcolumntype{Y}{>{\centering\arraybackslash}X}
\usepackage{balance}
\usepackage{subcaption}
\usepackage{hyperref}
\usepackage{url}
\usepackage{xspace}
\usepackage{tcolorbox}
\usepackage{enumitem}
\usepackage{multirow}
\usetikzlibrary{shapes}
\usetikzlibrary{patterns, fit}
\usepackage{adjustbox}
\usepackage{pifont}
\setminted[java]{
xleftmargin=15pt,
framesep=2mm,
linenos=true,
breaklines=true,
tabsize=2,
encoding=utf8,
frame=none,
fontsize=\scriptsize,
escapeinside=|| 
}

\newcommand{\highlight}[1]{\begin{tcolorbox}[leftrule=0mm,rightrule=0mm,toprule=0mm,bottomrule=0mm,left=0pt,right=0pt,top=0pt,bottom=0pt]
#1
\end{tcolorbox}
}

\makeatletter
\newcommand{\ostar}{\mathbin{\mathpalette\make@circled\star}}
\newcommand{\make@circled}[2]{%
  \ooalign{$\m@th#1\smallbigcirc{#1}$\cr\hidewidth$\m@th#1#2$\hidewidth\cr}%
}
\newcommand{\smallbigcirc}[1]{%
  \vcenter{\hbox{\scalebox{0.97778}{$\m@th#1\bigcirc$}}}%
}
\newbox\dottedarrow@box
\setbox\dottedarrow@box\hbox
  {%
    \begin{tikzpicture}
      \draw[dotted,->] (0,0) -- (1.5em,0);
    \end{tikzpicture}%
  }
\newcommand*\dottedarrow
  {\relax\ifmmode\expandafter\dottedarrow@m\else\expandafter\dottedarrow@t\fi}
\newcommand*\dottedarrow@t[1][1.5em]
  {\resizebox{#1}{!}{\raisebox{.5ex}{\usebox\dottedarrow@box}}}
\newcommand*\dottedarrow@m[1][]
  {%
    \if\relax\detokenize{#1}\relax
      \mathchoice
        {\dottedarrow@t}
        {\dottedarrow@t}
        {\dottedarrow@t[1.1em]}
        {\dottedarrow@t[0.9em]}%
    \else
      \dottedarrow@t[#1]%
    \fi
  }
\makeatother

\newcolumntype{g}{>{\columncolor{gray!30}}l}
\newcolumntype{h}{>{\columncolor{gray!30}}c}

\newcommand{\jucify}[0]{\textsc{JuCify}\xspace}
\newcommand{\flowdroid}[0]{\textsc{FlowDroid}\xspace}
\newcommand{\androzoo}[0]{\textsc{AndroZoo}\xspace}

\newcommand{\soot}[0]{\textsc{Soot}\xspace}
\newcommand{\angr}[0]{\textsc{Angr}\xspace}
\newcommand{\nativediscloser}[0]{\textsc{NativeDiscloser}\xspace}
\newcommand{\androguard}[0]{\textsc{AndroGuard}\xspace}

\newcommand{\droidbench}[0]{\textsc{DroidBench}\xspace}
\newcommand{\jimple}[0]{\textsc{Jimple}\xspace}

\newcommand{\nativescanner}[0]{\textsc{Native-Scanner}\xspace}
\newcommand{\doop}[0]{\textsc{DOOP}\xspace}
\newcommand{\etal}{\textit{et al.}\xspace}

\def\BibTeX{{\rm B\kern-.05em{\sc i\kern-.025em b}\kern-.08em
    T\kern-.1667em\lower.7ex\hbox{E}\kern-.125emX}}
    
\begin{document}

\title{JuCify: A Step Towards Android Code Unification for Enhanced Static Analysis}

\author{Jordan Samhi$^1$, Jun Gao$^1$, Nadia Daoudi$^1$, Pierre Graux$^{1,3}$, Henri Hoyez$^4$, Xiaoyu Sun$^2$, Kevin Allix$^1$, Tegawendé F. Bissyandé$^1$, Jacques Klein$^1$}
\affiliation{%
  \institution{$^1$SnT, University of Luxembourg, Luxembourg, firstname.lastname@uni.lu \\ 
  $^2$Monash University, Australia, firstname.lastname@monash.edu \\
  $^3$Univ. Lille, CNRS, Centrale Lille, UMR 9189 CRIStAL, F-59000 Lille, France, firstname.lastname@univ-lille.fr \\
  $^4$Technische Universität Kaiserslautern, Germany, firstname.lastname@sms-group.com \\}
  \country{}
}

\pagestyle{plain}

\begin{abstract}
Native code is now commonplace within Android app packages where it co-exists and interacts with Dex bytecode through the Java Native Interface to deliver rich app functionalities. Yet, state-of-the-art static analysis approaches have mostly overlooked the presence of such native code, which, however, may implement some key sensitive, or even malicious, parts of the app behavior. This limitation of the state of the art is a severe threat to validity in a large range of static analyses that do not have a complete view of the executable code in apps. To address this issue, we propose a new advance in the ambitious research direction of building a unified model of all code in Android apps. The \jucify approach presented in this paper is a significant step towards such a model, where we extract and merge call graphs of native code and bytecode to make the final model readily-usable by a common Android analysis framework: in our implementation, \jucify builds on the Soot internal intermediate representation. We performed empirical investigations to highlight how, without the unified model, a significant amount of Java methods called from the native code are ``unreachable'' in apps' call-graphs, both in goodware and malware. Using \jucify, we were able to enable static analyzers to reveal cases where malware relied on native code to hide invocation of payment library code or of other sensitive code in the Android framework.
Additionally, \jucify's model enables state-of-the-art tools to achieve better precision and recall in detecting data leaks through native code.
Finally, we show that by using \jucify we can find sensitive data leaks that pass through native code.
\end{abstract}

\maketitle

\section{Introduction}
\label{sec:introduction}
Android app analysis has been one of the most active themes of software engineering research in the last decade. Static analysis research, in particular, has produced a variety of approaches and tools that are leveraged in a variety of tasks, including bug detection, security property checking, malware detection, or empirical studies. The widely-used state-of-the-art approaches, such as FlowDroid~\cite{arzt2014flowdroid}, develop analyses that focus on the Dex bytecode in apps. Unfortunately, recent studies~\cite{afonso2016going,10.1145/3243734.3243835,lindorfer2014andrubis,6903578,tam2015copperdroid} have shown that malware authors often build on native code to hide their malicious operations (e.g., private data leak) or to implement sandbox evasion~\cite{evasion_techniques}.

The need to account for native code within Android apps is becoming urgent as the usage of native code is growing within both benign and malicious apps. 
Our empirical investigation on apps from the AndroZoo~\cite{Allix:2016:ACM:2901739.2903508} repository reveals that, in 2019, up to \num{62.9}\% of collected apps included native code within their packages. Yet, native code is scarcely considered in app security vetting~\cite{10.1145/3243734.3243835,ALAM2017230}. In the majority of static~\cite{7792435,7546513,doi:10.1155/2015/479174,papp2017towards,zhao2020automatic,arzt2014flowdroid,li2015iccta}, dynamic~\cite{petsas2014rage,bayer2006dynamic,zheng2012smartdroid} and machine learning based techniques~\cite{6735264,6298824}, native code is overlooked since it presents several challenges.

When researchers propose techniques to address native code such as with \emph{JN-SAF}~\cite{10.1145/3243734.3243835}, \emph{DroidNative}~\cite{ALAM2017230}, \emph{NativeGuard}~\cite{10.1145/2627393.2627396}, \emph{TaintArt}~\cite{10.1145/2976749.2978343} and others~\cite{9286029, 6903578,afonso2016going,george2020native}, the integrated analyses  (e.g., for taint tracking, native entry-point detection and machine learning feature extraction) are generally ad-hoc. Indeed, these works develop custom techniques to bridge native code and bytecode, typically by combining results of separate analyses of bytecode and native code. Therefore,
they do not yield an explicit unified model of the app to which generic analyses can be applied to explore bytecode and native code altogether. 

Our work aims to fill the gap in whole-app analysis by researching means to build a unified model of Android code. We propose \jucify, a  step toward building a framework that breaks bytecode-native boundaries for Android apps and therefore copes with a common limitation of static approaches in the literature. 
To the best of our knowledge, \jucify is the first approach that targets the unification of Android bytecode and native code into a unified model and is instantiated in a standard representation~\cite{LI201767}.
We target the \emph{Jimple}~\cite{vallee1998jimple} Intermediate Representation as support for \jucify unified model. Jimple is the internal representation in the widely-used \soot framework and is indeed the representation that is considered in a large body of static analysis works~\cite{LI201767}. By supporting Jimple, \jucify provides the opportunity for several analyses in the literature to readily account for native code.

{\bf This paper.} 
\jucify is a multi-step static analysis approach that we implement as a framework for generating a unified model of apps taking into account native code. It \ding{182} relies on symbolic execution to retrieve invocations between both the Dex bytecode and the native worlds, \ding{183} pre-computes native call-graph, \ding{184} merges Dex bytecode and native call-graphs, and \ding{185} populates newly generated functions with heuristic-based defined Jimple statements using code instrumentation.

The main contributions of our work are as follows:
\begin{itemize}[noitemsep,topsep=0pt]
    \item We propose \jucify, an approach to build a unified model of Android app code for enabling enhanced static analyses. We have implemented \jucify to produce the Jimple code that unifies bytecode and native code within an app package;
    \item We conduct an assessment of the \jucify yielded model. We show that \jucify can significantly enhance Android apps' call-graphs. \jucify connects previously unreachable methods in Android apps call-graphs;
    \item We evaluate the unified model of app code in the task of data flow tracking.
    We show that \jucify can significantly boost the precision of the state-of-the-art \flowdroid, from 0\% to 82\% and its recall from 0\% to 100\% on a new benchmark targeting bytecode-native data flow tracking;
    \item We evaluate \jucify on a set of real-world Android apps and show that it can augment existing analysers, enabling them to reveal sensitive data leaks that pass through the native code which were previously undetectable.
    \item We release our open-source prototype \jucify to the community as well as all the artifacts used in our study at:
    \begin{center}
       \url{https://github.com/JordanSamhi/JuCify} 
    \end{center}
\end{itemize}

The remainder of this paper is organized as follows. We first introduce background notions and motivate our work in Section~\ref{sec:background}. 
In Section~\ref{sec:approach}, we present our \jucify approach.
We evaluate \jucify in Section~\ref{sec:evaluation}.
In Sections~\ref{sec:limitations} and ~\ref{sec:threat_to_validity}, we present the limitations and the threats to validity of the current state of our approach.
Finally, we overview the related work in Section~\ref{sec:related_work} and conclude in Section~\ref{sec:conclusion}.
\section{Background \& Motivation}
\label{sec:background}

Java and Kotlin are the two mainstream programming languages that support the development of Android apps. Their programs are compiled into Dex bytecode and included within app packages (in the form of DEX files).
Nevertheless, thanks to \emph{Java Native Interface}~\cite{jni}, native code functionalities are accessible in Android apps. They come in binary (e.g., \texttt{.so} share library) files compiled from input programs written in C/C++ for instance.

\subsection{Java Native Interface (JNI)}
\label{sec:background:jni}

JNI is an implementation of
the \emph{Foreign Function Interface} (FFI)~\cite{10.1145/1065010.1065019} mechanism that allows programs written in a given language to invoke subroutines written in another language.
JNI allows both Java to native and native to Java invocations.

\subsubsection{Java to native code}
\label{sec:background:jni:javatonative}
Listing~\ref{code:motivation} presents an example where JNI capabilities are used to call a native function (here written in C$++$) from Java.
First, a relevant Java method is defined with the keyword \texttt{native} (line 4). We will refer to it as a \emph{Java native method}. 
Then, its corresponding native function is registered to set up the mapping between them.
Such a registration can be:

\noindent
\textbf{Static -} the native function definition follows a naming convention based on specific JNI macros. For example, the Java native method \texttt{native\-Get\-Imei} (line~4) corresponds to a native function named \texttt{Java\_com\_\-example\_\-nativeGetImei} in C$++$~(line~16). 

\noindent
\textbf{Dynamic -} developers can arbitrarily name their native functions (in C$++$) as shown in Listing~\ref{code:dynamic_registration} (lines 10-13), but must inform JNI about how to map them with Java native methods. Thus, developers 
\ding{182} first map Java native methods to their counterpart native functions by using specific \texttt{JNINative\-Method} structures (lines 14-16 in Listing~\ref{code:dynamic_registration});
\ding{183} overload a specific JNI Interface function~\cite{oracle-jni}, \texttt{JNI\_OnLoad}, to register the mapping (lines 17-24 in Listing~\ref{code:dynamic_registration});
and \ding{184} invoke \texttt{Register\-Natives} in \texttt{JNI\_OnLoad} which will be called by Android VM (line 22 in Listing~\ref{code:dynamic_registration}).

\subsubsection{Native to Java}
\label{sec:background:jni:nativetojava}

With JNI, developers can create and manipulate Java objects within the native code (e.g., written in C$++$).
The fields and methods of Java objects are also accessible from the native code and can be invoked using specific JNI \textit{Interface} functions.
Eventually, likewise Java reflection~\cite{forman2004java}, i.e., using strings to get methods and classes, the developer can invoke the Java methods (e.g., lines 17-19 in Listing~\ref{code:motivation}).

\highlight{
Note that Listings~\ref{code:motivation} and~\ref{code:dynamic_registration}  illustrate the interaction between Java and C$++$. However, \jucify, the approach proposed in this paper, works at the apk level.
Therefore, the invocations are between bytecode and compiled native code.
}

\begin{listing}
\begin{minted}{java}
/*** JAVA WORLD ***/
public class MainActivity extends Activity {
  static {System.loadLibrary("native-lib");}
  public native String nativeGetImei(TelephonyManager tm);
  @Override
  protected void onCreate(Bundle savedInstanceState) {
    super.onCreate(savedInstanceState);
    TelephonyManager tm =  (TelephonyManager) getSystemService("phone");
    String imei = nativeGetImei(tm);
    Log.d("IMEI",  "" + imei);
  }
  public void malicious() {/* malicious code */}
}
/*** C++ WORLD ***/
JNIEXPORT jstring JNICALL
Java_MainActivity_nativeGetImei(JNIEnv *env, jobject thiz, jobject tm) {
  jclass c = (*env).GetObjectClass(tm);
  jmethodID m = (*env).GetMethodID(c, "getImei", "()Ljava/lang/String;");
  jstring i = (jstring)(*env).CallObjectMethod(tm, m);
  c = (*env).GetObjectClass(thiz);
  m = (*env).GetMethodID(c, "malicious", "()V");
  (*env).CallObjectMethod(thiz, m);
  return i;
}
\end{minted}
    \caption{Code illustrating how an app can trigger native code. (Methods and code are simplified for convenience)}
    \label{code:motivation}
\end{listing}

\begin{listing}
\begin{minted}{java}
/*** JAVA WORLD ***/
public class MainActivity extends Activity {
    static {System.loadLibrary("native-lib");}
    public native String nativeMethod();
    @Override
    protected void onCreate(Bundle b) {nativeMethod();}
}
/*** C++ WORLD ***/
JNIEXPORT jstring JNICALL
jstring arbitrary_name(JNIEnv *e, jobject thiz) {
    std::string str = "str";
    return e->NewStringUTF(str.c_str());
}
static const JNINativeMethod m[] = {
    {"nativeMethod", "()Ljava/lang/String;", (jstring*)arbitrary_name}
};
JNIEXPORT jint JNI_OnLoad(JavaVM* vm, void* reserved){
    JNIEnv* e = NULL;
    if(vm->GetEnv((void**)&e, JNI_VERSION_1_4) != JNI_OK){return -1;}
    jclass c = e->FindClass("com/example/MainActivity");
    if (!c){return -1;}
    if(e->RegisterNatives(c, m, sizeof(m)/sizeof(m[0]))){return -1;}
    return 1;
}
\end{minted}
    \caption{Dynamic native function registration example. (Methods and code are simplified for convenience)}
    \label{code:dynamic_registration}
\end{listing}
\subsection{Motivating Example}
\label{sec:motivation}

Binary static code analysis is in itself a challenge~\cite{10.1145/2931037.2931047} since the compiled code is hard to represent for appropriate investigation~\cite{1281665}. 

Although current state-of-the-art Android static code analysis approaches are sophisticated~\cite{arzt2014flowdroid,li2015iccta,samhi2020raicc,10.1145/2660267.2660357,6394931}, most of the time they overlook native code, with only a few of them considering it~\cite{10.1145/3243734.3243835,9286029}.

With a simple example illustrated in Listing~\ref{code:motivation}, we make the case that native code should be considered in static analysis approaches.

\textbf{First,} in the {\em onCreate()} method of the main \texttt{Activity}, 
a \texttt{String} is retrieved on line 9 from the method \emph{native\-Get\-Imei}, then this \texttt{String} is used as a parameter to the method \emph{Log.d()}.
From the point of view of taint tracking, there is a flow from the \emph{getImei()} method (source) to the \emph{Log.d()} method (sink).
However, most state-of-the-art approaches will miss this flow due to technical limitations since the method \emph{nativeGetImei} is not analyzed. 
Therefore the variable {\tt imei} is not tainted, and the flow is not detected.

\textbf{Second,} the method \emph{malicious()} (line 12) is never called in the Java code, thus, it will not appear in the call-graph since it is considered as \emph{unreachable}. Hence it will not be analyzed, causing existing tools to fail to detect potential malicious code in the method.

\begin{figure}[ht]
    \centering
    \begin{adjustbox}{width=.8\columnwidth,center}
    \begin{tikzpicture}
    \tikzstyle{arrowStyle}=[-latex]
    \tikzset{node/.style={ellipse,align=center,scale=.8}}

    \path
    
    (0,0) node[node,draw,fill=green!10] (oncreate2) {\small  onCreate}
    ++(-1,0.7) node[node,draw,fill=green!10] (d2) {\small d}
    ++(1.5,.2) node[node,draw,fill=green!10] (native2) {\small nativeGetImei}
    ++(-3,-1) node[node,draw,fill=green!10] (getss2) {\small get\-System\-Service}
    ++(4,0) node[node,draw,fill=red!10] (getImei) {\small getImei}
    ++(2,1) node[node,draw,fill=red!10] (getObjectClass) {\small GetObjectClass}
    ++(-.5,1) node[node,draw,fill=red!10] (getMethodId) {\small GetMethodId}
    ++(-3,0) node[node,draw,fill=red!10] (callobjectMethod) {\small CallObjectMethod}
    ++(3.2,-1.8) node[node,draw,fill=red!10] (malicious) {\small malicious}
    ;
    
    \draw[->,>=latex] (oncreate2) -- (d2);
    \draw[->,>=latex] (oncreate2) -- (native2);
    \draw[->,>=latex] (oncreate2) -- (getss2);
    \draw[->,>=latex] (native2) -- (getImei);
    \draw[->,>=latex] (native2) -- (getObjectClass);
    \draw[->,>=latex] (native2) -- (getMethodId);
    \draw[->,>=latex] (native2) -- (callobjectMethod);
    \draw[->,>=latex] (native2) -- (malicious);
    
    \draw[->,>=latex] (oncreate2) to[out=220,in=270] (0,-.45) to[out=70,in=270] (oncreate2);
    
\end{tikzpicture}
    \end{adjustbox}
    \caption{Unified call-graph representation for the code in Listing~\ref{code:motivation}: \textcolor{green}{Green} nodes represent reachable nodes of existing static approaches, while \textcolor{red}{Red} ones represent the nodes unreachable with most of the existing static approaches}
    \label{fig:motivation}
\end{figure}
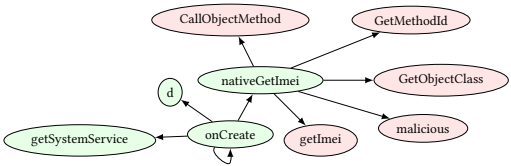

Let us consider Figure~\ref{fig:motivation}, which presents the expected call-graph of this example. The current state-of-the-art approaches, such as ~\cite{arzt2014flowdroid,li2015iccta,fratantonio2016triggerscope,10.1145/2660267.2660357,li2016droidra}, generally analyze the green nodes which are reachable from an entry point. 
However, the red nodes will only be considered by approaches able to analyze the native code.
Approaches trying to overcome the challenge of native code analysis in Android apps, already exist (e.g., ~\cite{9286029,10.1145/3243734.3243835,6903578,rasthofer2016harvesting}). 
However, they focus on specific analyses and propose custom solutions to bridge bytecode and native code. 
In contrast, in this paper, we aim at offering an explicit unified model of Android apps to which generic analysis could be applied to explore altogether bytecode and native code. 

\section{Approach}
\label{sec:approach}

\begin{figure*}[ht]
    {\begin{center}CG = Call-Graph, \textcolor{red!30}{\Huge $\bullet$} \textcolor{blue!30}{\Huge $\bullet$} = Native CG Node, \textcolor{green!30}{\Huge $\bullet$} = Bytecode CG Node, \textcolor{yellow!60}{\Huge $\bullet$} = Previously Unreachable Bytecode CG Node\end{center}} \medskip
    \medskip
    \centering
        \input{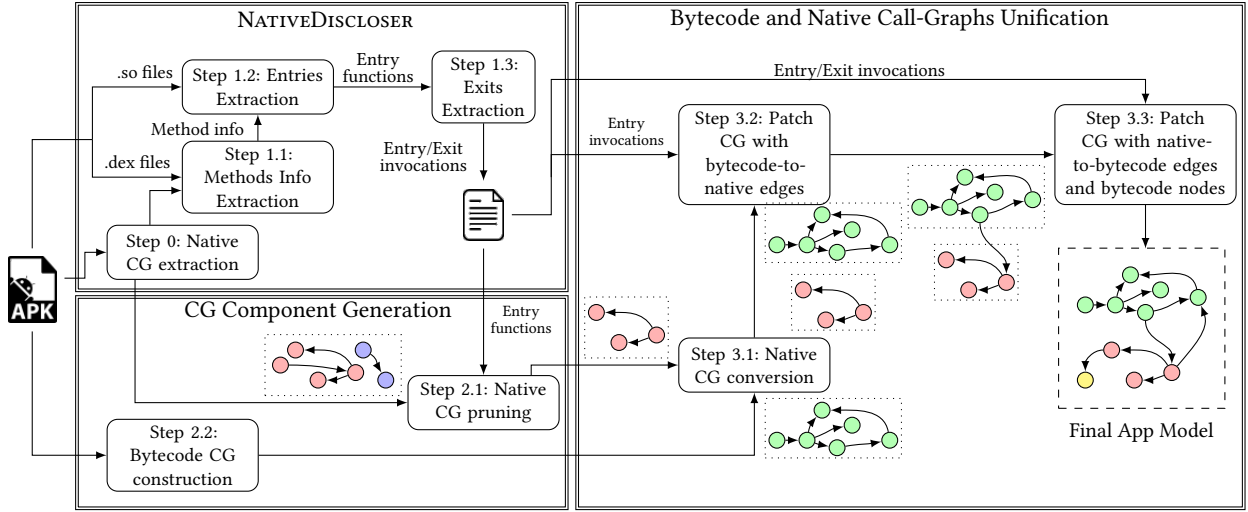}
    \caption{Overview of the \jucify Approach from the Angle of Call-Graph Construction}
    \label{fig:overview}
\end{figure*}

For a given Android app, \jucify aims to unify its Dex bytecode and native code into a unified model and instantiate this model in the Jimple representation (i.e., the intermediate representation of the popular Soot framework).
In this section, we will first detail the overall \jucify conceptual approach, and then we will briefly present how we instrument the app to approximate the native behavior. 
However, due to space constraints, we will not present all technical details related to Jimple. We invite the interested reader to consider all our publicly-shared artifacts on the project Github repository\footnote{\scriptsize \url{https://github.com/JordanSamhi/JuCify}}. \jucify implementation is fully open-sourced.

\subsection{Call Graph as Unified Preliminary Model}\label{sec:callGraph}
To explain the overall functioning of \jucify, we will restrict our explanations to the notion of Call Graph (CG). A CG can be defined as $CG = (V, E)$, where $V$ is a set of vertices representing functions, and $E \subseteq \lbrace(u,v) \; \vert \; u,v \in V\rbrace$ is a set of edges such as $\forall (u,v) \in E$, there is a call from $u$ to $v$ in the program. 

\jucify is a multi-step static analysis framework whose overall architecture is depicted in Figure~\ref{fig:overview}.
First, a submodule called \nativediscloser constructs the native callgraph and extracts the mutual invocations between bytecode and native code.
Then, native callgraph is pruned and prepared to be \soot-compliant before being merged with the bytecode callgraph.
Eventually, both callgraphs are unified thanks to information related to the bytecode-native method invocations.
In the following with give more details about the different steps of our approach.

\noindent
\textbf{Step 0: Native Call Graph Construction}

Native program call-graph construction is not trivial~\cite{10.1145/1127577.1127590}.
In fact, a large body of work tackled this problem and proposed several solutions to find function boundaries~\cite{10.1145/1127577.1127590,kinable2011malware,10.1145/279310.279314}.
In this work, the call-graphs native libraries in Android apps are generated by \angr~\cite{angr}, a well-known binary analysis framework, which is wrapped into our submodule \nativediscloser.

\noindent
\textbf{Step 1: Bytecode-Native Code Invocations Extraction}

This step is performed over 4 sub steps:
\ding{182} Retrieve bytecode methods information;  
\ding{183} Extract entry method invocations (i.e., bytecode to native);
\ding{184} Track native function calls and extract exit method invocations.

{\em Step 1.1: Methods info extraction} is a straightforward task that extracts information of bytecode methods, such as the class of a method, method signature. 
This step aims to complete the signature information required to perform the method invocations extraction task for statically registered functions. 
We perform this task by relying on \androguard~\cite{androguard}.

{\em Step 1.2: Entry method invocations extraction:}
An entry method invocation is a native method invocation from the bytecode (i.e., a bytecode-to-native "link"). 
As described in Section~\ref{sec:background:jni}, for such an invocation, we need to match a "Java native method" (i.e., a method declared in Java with the \texttt{native} keyword, also called \emph{entry method}) and an \emph{entry function} (i.e., the counterpart native function).
To perform this task, we have to take care of both static and dynamic registrations.
The statically registered functions can be easily spotted via their naming conventions.
However, as dynamic registration relies on JNI \textit{interface} function
calls, more sophisticated techniques are required. In our case, we rely on symbolic execution.

From a more technical point of view,  
\nativediscloser takes as input the library (i.e., \texttt{.so}) files of an apk and the method information from the previous step.
It first scans the symbol table of each binary to search for (1) statically registered native functions and (2) 
the \texttt{JNI\_OnLoad} function for the case of dynamically registered functions.
Then, if \texttt{JNI\_OnLoad} exists, this function is symbolically executed to further detect dynamically registered native functions.

For symbolic execution, \nativediscloser relies on \angr~\cite{angr}.

{\em Step 1.3: Exit method invocations extraction:} 
We are looking for the invocations of a bytecode method from the native code. 
We call \emph{exit method} this bytecode method. 
In Section~\ref{sec:background:jni:nativetojava}, we explained that this exit method is called by invoking certain JNI \textit{Interface} functions in a chained manner.
Collecting information related to this chain of JNI function invocations is challenging.

In practice, to overcome this challenge, \nativediscloser executes all the entry functions acquired from step 1.2 symbolically to search for the exit method invocations and set up the relation mapping between entry and exit method invocations.

Furthermore, exit methods could be invoked deep down in a native function chain.
However, the symbolic execution is not aware of the boundaries between native functions.
Hence, we implemented a tracking mechanism during the search of exit methods.
We rely on the starting address of each native function obtained from the native call-graph to maintain a stack of native functions and push a new function into the stack when its starting address is reached.
Popping a function from the stack is triggered by the arrival of the return addresses of native functions, which can be obtained from a certain register or memory location based on architecture specifications (e.g., link register \textit{LR} for \textit{ARM}) during entering a native function.
This allows us to know from which native function an exit method invocation occurs.

\noindent
\textbf{Step 2: CG Components Generation}

{\em Step 2.1: Native CG pruning.} Since in .so libraries not all the functions are necessarily called in an app, we rely on a strategy to only keep relevant callgraph parts.
To do so, we prune the obtained native call-graphs constructed in Step 0 with the help of the entry functions passed in from Step 1.
We only keep the sub-graphs starting from the entry functions (with all successor nodes) since the remaining parts will not be reachable from the bytecode.

{\em Step 2.2: Bytecode CG construction.} Our approach also requires the bytecode call-graph. 
For this purpose, we use \flowdroid~\cite{arzt2014flowdroid} (itself based on \soot~\cite{vallee2010soot}) which leverages an advanced modeling of app components' life-cycle.

\noindent
\textbf{Step 3: Bytecode and native call-graphs unification}

{\em Step 3.1: Native CG conversion.}
In practice, the target is to load both native and bytecode call-graphs in \soot.
Although this is straightforward for the bytecode call-graph, the native call-graph requires a conversion step to fit with \soot technical constraints.
Once loaded, the sets of nodes and edges of both call-graphs are merged, but the call-graphs are not yet connected together. 

{\em Step 3.2: Patch CG with bytecode-to-native edges.}
Then, according to the entry invocations obtained from Step~1.2, edges between entry methods (in bytecode) and their counterpart entry functions (in native code) are added.

{\em Step 3.3: Patch CG with native-to-bytecode edges and bytecode nodes.}
Finally, with the information of exit invocations and the relations with entry invocations from Step~1.3, edges between native functions to exit methods are added.
This step allows uncovering previously unreachable bytecode callgraph nodes.

\subsection{From CG to Jimple for a Unified Model}
\label{sec:cg_to_jimple}

A call-graph is a useful model, but it is still limited because it does not contain enough information to perform static analysis (e.g., data flow analysis).
Indeed, important information such as the statements present in each method is missing (i.e., the control flow graph (CFG)). 
A tool such as \flowdroid provides the CFG for each bytecode method where the method behavior is represented with Jimple statements.  
We will now explain how \jucify adds Jimple statements in specific native functions in a best-effort mode.
After this step, for a given APK, we obtain the Jimple representation of the apk with both bytecode and native code unified.  

\noindent
\textbf{Native functions generation:} 
\jucify relies on a \emph{DummyBinaryClass} whose purpose is to incorporate any newly imported native function in the \soot representation.
For each native function in the native call-graph, \jucify generates a new method in the \emph{DummyBinaryClass} with appropriate signatures.

\noindent
\textbf{Bytecode method statements instrumentation:} 
\jucify generates bytecode-to-native call-graph edges.
It also has to replace the initial call to the native method at the statement level with a call to the newly generated native function.
\jucify takes care of the returned value and the parameters to not fool any analysis based on the new built model.

\noindent
\textbf{Native function statements generation:}
There is no bijection between native code and \jimple code~\cite{vallee1998jimple}.
Moreover, bytecode and native code manipulate different notions (e.g., pointers) that cannot be translated directly.
Therefore, we have to use heuristics based on the information at our disposal to put a first step toward reconstructing native function behavior.

Let us consider a native function named \texttt{foo()} containing at least one invocation to a bytecode method $m$.
As explained in Section~\ref{sec:callGraph}, the first step of \jucify aims to collect information about bytecode methods (full signature). 
Thanks to this, we can approximate the parameters used by $m$ as well as its return values.

More specifically, in Listing~\ref{code:native_to_java}, we detail the steps \jucify implements to populate the native function \texttt{foo()} that calls a bytecode method $m$. Let consider $m$ is defined in a Java class named \texttt{MyClass}.
In line 1, \jucify starts with the empty method \texttt{foo()}. Then:

\noindent
\textbf{Step 1 in Listing~\ref{code:native_to_java}:} If the bytecode method $m$ should return a value, \jucify generates a new local variable with the same type as the method's return type (line 4).

\noindent
\textbf{Step 2 in Listing~\ref{code:native_to_java}:} \jucify generates the declaration of a variable of type \texttt{MyClass}, the class in which $m$ is defined (line 8).
In line 9, \jucify creates a new \texttt{MyClass} instance (if there is not one usable as a base for the bytecode call).

\noindent
\textbf{Step 3 in Listing~\ref{code:native_to_java}:} Regarding the parameters that should be used for the invocation of $m$, \jucify scans \texttt{foo()} for local variables and parameters whose types match the types of the parameters of $m$.
If, for a given type, no local variable, nor parameter of \texttt{foo()} is found, \jucify generates one (e.g., line 15).
Then, it generates all the permutations of these variables with a given length (i.e., the number of parameter of $m$) and retains only those matching the types' order of the parameters of $m$ ($(i1, s)$, and $(i2, s)$ in Listing~\ref{code:native_to_java}).
Each retained permutation corresponds to a possible call to the bytecode method in the native function as an over-approximation.
Nevertheless, these calls cannot be generated sequentially since they correspond to different realities.
Hence, we rely on opaque predicates (\emph{if} statements whose predicate cannot be evaluated statically) so that each control flow path is considered identically (lines 16-17).

\noindent
\textbf{Step 4 in Listing~\ref{code:native_to_java}:} If the native function returns a value (from the signature of \texttt{foo()}), \jucify should generate return statements.
To do so, it operates as for $m$. It relies on opaque predicates.
Indeed, first, \jucify scans the body of the current native function to find any local variable corresponding to the type of the return value (even those newly generated local variables that could be returned).
If no variable is found, \jucify generates such a variable.
Else, for each of found local variables, \jucify generates return statements with opaque predicates so that each path can be equally considered (lines 26-27 in Listing~\ref{code:native_to_java}).

\begin{listing}
    \medskip
\begin{minted}{java}
public boolean foo(int i1, int i2, boolean b1){}
// STEP 1
public boolean foo(int i1, int i2, boolean b1){
    boolean b2;
}
// STEP 2
public boolean foo(int i1, int i2, boolean b1){
    boolean b2; MyClass jc;
    jc = new MyClass();
}
// STEP 3
public boolean foo(int i1, int i2, boolean b1){
    boolean b2; MyClass jc; String s;
    jc = new MyClass();
    s = new String();
    if(opaque_predicate) {b2 = jc.m(i1, s);}
    else if(opaque_predicate) {b2 = jc.m(i2, s);}
}
// STEP 4
public boolean foo(int i1, int i2, boolean b1){
    boolean b2; MyClass jc; String s;
    jc = new MyClass();
    s = new String();
    if(opaque_predicate) {b2 = jc.m(i1, s);}
    else if(opaque_predicate) {b2 = jc.m(i2, s);}
    if(opaque_predicate) {return b1;}
    else if(opaque_predicate) {return b2;}
}
\end{minted}
    \caption{\jucify's process to populate native functions}
    \label{code:native_to_java}
\end{listing}

Finally, \jucify yields a unified model of Android apps on which analysts can perform any static analysis.
\section{Evaluation}
\label{sec:evaluation}

We investigate the following research questions to assess the importance of our contributions:

\begin{description}
    \item[RQ1:] What is the proportion and evolution of native code usage in both real-world benign and malicious apps?
    \item[RQ2:] To what extent our bytecode-native invocation extraction step (named \nativediscloser) yields better results than the state-of-the-art ?
    \item[RQ3:] Can \jucify boost existing static data flow analyzers?
    \item[RQ4:] How does \jucify behave in the wild? We address this question both at the quantitative and qualitative levels:
    \begin{itemize}
        \item \textbf{RQ4.a:} To what extent can \jucify augment apps' call-graphs and reveal previously unreachable Java methods?
        \item \textbf{RQ4.b:} Can \jucify reveal previously unreachable data leaks that pass through native code in real-world apps?
    \end{itemize}
\end{description}

\subsection{RQ1: Native code usage in the wild}
\label{sec:evaluation:native_usage}

This section presents general statistics about the usage of native code in both benign and malicious Android apps. 
We also perform an evolutionary study of this usage. 

\textbf{Dataset:} We rely on the \androzoo repository~\cite{Allix:2016:ACM:2901739.2903508}  to build 
\ding{182} a dataset of \num{2641194} benign apps (where we consider an app as benign if no Antivirus in VirusTotal~\cite{total2012virustotal} has flagged it - score 0);
and \ding{183} a dataset of  \num{174342} malicious apps (where we consider an app as malicious when at least 10 Antivirus engines in VirusTotal have flagged it). 
Both datasets contain all the apps from 2015 to 2020 that we were able to collect from \androzoo with the mentioned VirusTotal constraints.

\textbf{Empirical study:} Android programming with the Native Development Kit (NDK) suggests developers to integrate native libraries (i.e., {\tt .so} files) whose code can be invoked from the Java world. 
Therefore, to study the extent of native code usage in Android apps, as a preliminary study, for each app, we check if it contains at least one .so file in its APK file.
However, since native libraries can be present in apps but never used, we also check for each app if Java native methods (cf. Section~\ref{sec:background:jni}) are declared in the bytecode.

\begin{table}[ht!]
      \begin{adjustbox}{width=\columnwidth,center}
        \begin{tabular}{l|c|c|c||c|c|c}
            & \multicolumn{3}{c||}{\textbf{Goodware}} & \multicolumn{3}{c}{\textbf{Malware}} \\
            \hline
            & \# Apps & w/ .so files & w/ native methods & \# Apps  & w/ .so files & w/ native methods \\
            \hline
            2015 & \num{632279} & \num{220934} (\num{34.9}\%) & \num{216329} (\num{34.2}\%) & \num{89542} & \num{65221} (\num{72.8}\%) & \num{63275} (\num{70.7}\%) \\ 
            2016 & \num{1103899} & \num{405209} (\num{36.7}\%) & \num{404357} (\num{36.6}\%) & \num{48358} & \num{35601} (\num{73.6}\%) & \num{34240} (\num{70.8}\%) \\ 
            2017 & \num{277690} & \num{143463} (\num{51.7}\%) & \num{143183} (\num{51.6}\%) & \num{15141} & \num{8742} (\num{57.7}\%) & \num{8539} (\num{56.4}\%) \\ 
            2018 & \num{304746} & \num{191491} (\num{62.8}\%) & \num{184447} (\num{60.5}\%) & \num{10890} & \num{8415} (\num{77.3}\%) & \num{8018} (\num{73.6}\%) \\ 
            2019 & \num{179309} & \num{113433} (\num{63.3}\%) & \num{112873} (\num{62.9}\%) & \num{9773} & \num{8993} (\num{92.0}\%) & \num{8311} (\num{85.0}\%) \\ 
            2020 & \num{143271} & \num{81755} (\num{57.1}\%) & \num{81111} (\num{56.6}\%) & \num{638} & \num{446} (\num{69.9}\%) & \num{274} (\num{42.9}\%) \\ \hline
            Total & \num{2641194} & \num{1156285} (\num{44}\%) & \num{1142300} (\num{43}\%) & \num{174342} & \num{127418} (\num{73}\%) & \num{122657} (\num{70}\%) \\ 
        \end{tabular}
        \end{adjustbox}
    \caption{Number and proportion of Android apps that contain at least one "\texttt{.so} file" / "Java native method" (w/ = with).}
    \label{table:proportion_native_code}
\end{table}

\textbf{Results} of our empirical study are presented in Table~\ref{table:proportion_native_code}. 
They indicate that, overall, \num{1156285} benign apps (i.e., \num{44}\%) contain at least one .so file, and \num{1142300} (i.e., \num{43}\%) contain at least one Java native method declaration.
This means that \num{98.8}\% of apps with native libraries contain Java native method declaration in their bytecode.
Regarding malware, \num{127418} (i.e., \num{73}\%) of apps contain native libraries and \num{122657} (i.e., \num{70}\%) Java native method declarations.
Hence, \num{96.3}\% of malware with native libraries contain Java native method declarations. 
Overall, these results show that native code is, in proportion, more used in malicious apps. 

Regarding usage evolution in benign apps, the rate increases until 2018 to reach a plateau at around $60\%$. 
The trend regarding malware is much more erratic (with sharp decreases in 2017 and 2020).
However, for each year, malicious apps use significantly more native code than benign apps.

\highlight{
\textbf{RQ1 answer:}
Native code is definitely pervasive in Android apps.
While both benign and malicious code leverage native code, native invocations are substantially more common in malware (\num{70}\% vs. \num{43}\%).

These results indicate that ignoring native code is a serious threat to validity in Android static code analysis.
}

\subsection{RQ2: Bytecode-Native Invocation Extraction Comparison }
\label{sec:evaluation:comparison}

Identifying native-to-bytecode and bytecode-to-native code invocations are key steps towards code unification. 
Our objective is to estimate to what extent the corresponding building block in \jucify is effective against a benchmark and against the state of the art.

\textbf{Native to Bytecode:} Fourtounis \etal~\cite{george2020native} proposed an approach to detect exit invocations (i.e., native to bytecode invocations, c.f., Section~\ref{sec:callGraph}) in native code via binary scanning. 
Their tool named \nativescanner~\cite{nativeScanner}
has been developed as a plugin of a framework called \doop~\cite{doop}.
Briefly, their tool scans binary files for string constants that match Java method names and Java VM type signatures and follows their propagation.
In this way, they consider all matches as new entry points back to bytecode.

To compare our \nativediscloser with \nativescanner, we developed and released 16 benchmark apps.
All these apps are executable Android apps and have been tested on a \textit{Nexus} 5 phone with Android version 8.1.0.
We design these apps to cover different situations such as dynamic/static registration, chained invocations in native functions,
parameter passing via structures and classes, string accessing via arrays and function returns, string obfuscation, etc.
Table~\ref{tab:comparison} presents the results obtained with both tools.

\begin{table}[h!]
    {\begin{center}\scriptsize \textcolor{green}{TP} = True Positive, \textcolor{red}{FP} = False Positive, \textcolor{magenta}{FN} = False Negative\end{center}}
    \begin{adjustbox}{width=.8\columnwidth,center}
    \begin{tabularx}{\linewidth}{c|Y|Y|Y||Y|Y|Y}
        \hline
        \multirow{2}{*}{\textbf{Benchmark}} & \multicolumn{3}{c||}{\textbf{\scriptsize \nativescanner}} & \multicolumn{3}{c}{\textbf{\scriptsize \nativediscloser}} \\
        \cline{2-7}
        & \textcolor{green}{\scriptsize TP} & \textcolor{red}{\scriptsize FP} & \textcolor{magenta}{\scriptsize FN} & \textcolor{green}{\scriptsize TP} & \textcolor{red}{\scriptsize FP} & \textcolor{magenta}{\scriptsize FN} \\
        \hline
        \hline
        $bm_{1-5}$, $bm_{7}$, $bm_{10-12}$$^\dagger$  & 1 & 0 & 1 & 2 & 0 & 0\\
        \hline
        $bm_{6}$, $bm_{8}$ & 1 & 0 & 2 & 3 & 0 & 0 \\
        \hline
        $bm_{9}$ & 0 & 0 & 2 & 0 & 0 & 2 \\
        \hline
        $bm_{13}$ & 0 & 1 & 5 & 5 & 0 & 0 \\
        \hline
        $bm_{14}$ & 1 & 4 & 1 & 2 & 0 & 0 \\
        \hline
        $bm_{15}$, $bm_{16}$ & 1 & 0 & 1 & 1 & 0 & 1 \\
        \hline
        \hline
        Precision & \multicolumn{3}{r||}{73.68\%} & \multicolumn{3}{r}{100\%} \\
        \hline
        Recall & \multicolumn{3}{r||}{37.84\%} & \multicolumn{3}{r}{89.19\%} \\
        \hline
    \end{tabularx}
    \end{adjustbox}
    $^\dagger$ due to space limitation, we put together apps with same results. E.g., \nativediscloser detects 3 TP for each app $bm_6$ and $bm_8$.
    \caption{Comparison of Tools}
    \label{tab:comparison}
\end{table}

These results show that \nativescanner misses a high number of exit invocations.
We realized that \nativescanner seems not to consider Android framework APIs (the tool misses the API invocations in all benchmark apps). 
Note that \nativescanner is not specific to Android. This could explain why it does not consider Android APIs.
The tool is also challenged by constant string obfuscation (app $bm_9$), which is also the case for \nativediscloser. 
\textit{bm14} implements fake method string constants in the native part. 
For this app, we can observe the over-estimation of \nativescanner (i.e., a high number of false-positive) while \nativediscloser is not affected.
Finally, \nativediscloser also failed with string constants passing via arrays and function returns as implemented in \textit{bm15} and \textit{bm16} respectively.
Limitations of \angr could cause this in parsing pointer of pointers.
Overall, compared to \nativescanner, \nativediscloser obtains significantly higher precision and recall.

\textbf{Bytecode to native:} 
We were unable to compare \nativediscloser with \nativescanner.
Unlike our tool, \nativescanner does not investigate (1) bytecode to native entry invocations and (2) the relations mapping between entry and exit invocations.

Note, however, that on our benchmark of 16 apps, \nativediscloser yields 100\% precision in finding both the entry invocations and the entry-to-exit relations and achieves a recall of 95.59\% and 89.19\% respectively.

\highlight{
\textbf{RQ2 answer:}
Compared to the state-of-the-art \nativescanner, our
\nativediscloser extracts exit invocations with better precision and recall.
Besides, it can provide extra information, including entry invocations (i.e., bytecode to native invocations) and the relations with exit invocations, which is essential to generate comprehensive call-graphs.
}

\subsection{RQ3: Can \jucify boost static data flow analyzers?}
\label{sec:evaluation:data_flow}

In Section~\ref{sec:approach}, we described how \jucify could approximate the behavior of native functions based on the information retrieved from signatures, parameters, return type, and bytecode methods called from native code via JNI.
In this RQ, we check if this first step approximation helps perform advanced static analyses such as data leak detection on a well-defined benchmark. We will assess the capability of \jucify on real-world applications in RQ4.

The benchmark that we built for RQ3 contains 11 apps that we plan to integrate into \droidbench, an open test suite that contains hand-crafted Android apps to assess taint analyzers. 
Among these apps, 9 contain a flow going through the native world, and 2 do not contain any data flow (to detect potential false positives).
Then, we apply the state-of-the-art \flowdroid taint-analysis engine before and after applying \jucify in our benchmark apps, to show that \flowdroid can, likewise in~\cite{samhi2020raicc}, be \emph{boosted}.
\flowdroid detects paths from well-defined sources (e.g., \texttt{getDeviceId()}) and sinks (e.g., \texttt{sendTextMessage()}) methods in Android apps. 

\textbf{Benchmark construction:} We identified 4 cases on which we built our 11 benchmark apps to assess the ability of tools in detecting data leak via native code:
\begin{enumerate}
    \item[a)] Getter: Source in native code and sink in Java code
    \item[b)] Leaker: Source in Java code and sink in native code
    \item[c)] Proxy: Source in Java code and sink in Java code
    \item[d)] Delegation: Source in native code and sink in native code
\end{enumerate}

\begin{figure}[ht!]
    {\begin{center}\textcolor{red}{$\bullet$} = Tainted Information, \dottedarrow\ = Call Edge, \textcolor{red}{$\rightarrow$} = Taint Propagation, \textcolor{black}{$\bullet$} = Method entrypoint\end{center}}
    \begin{subfigure}{.49\linewidth}
        \centering
        \resizebox{.9\linewidth}{!}{
            \begin{tikzpicture}
    \tikzstyle{arrowStyle}=[-latex]
    \tikzset{node/.style={minimum height=0.9cm, rounded corners, text width=1.9cm,align=center,scale=.8}}

    \path
    (-1,-.5) node[node,draw] (nativecall) {  = native()}
    ++(0,-1) node[node,draw] (sink) {sink( )}
    ++(3,1) node[node,draw] (source) { = source( )}
    ++(0,-1) node[node,draw] (return) {return}
    ;
    
    \node[circle,draw=black, fill=black, inner sep=0pt,minimum size=3pt] (a) at (-1,.5) {};
    
    \node[circle,draw=red, fill=red, inner sep=0pt,minimum size=3pt] (a) at (1.3,-.5) {};
    \node[circle,draw=red, fill=red, inner sep=0pt,minimum size=3pt] (b) at (-1.65,-.5) {};
    \node[circle,draw=red, fill=red, inner sep=0pt,minimum size=3pt] (c) at (2.5,-1.5) {};
    \node[circle,draw=red, fill=red, inner sep=0pt,minimum size=3pt] (d) at (-.78,-1.5) {};
    
    \draw[->,>=latex,red] (a) to[out=350,in=130] (c);
    \draw[->,>=latex,red] (c) to[out=150,in=330] (b);
    \draw[->,>=latex,red] (b) to[out=270,in=90] (d);
    
    \draw[->,>=latex,black] (-1,.5) -- (nativecall);
    
    \draw[->,>=latex,black] (nativecall) -- (sink);
    \draw[->,>=latex,black] (source) -- (return);
    
    \draw[->,>=latex,densely dotted] (nativecall.north east) to[out=20,in=90] (2,.1);
    
    \node[fit={(-2,0) (-2,-2) (0,0) (0,-2)}, draw, dashed] (java) {};
    \node[fit={(1,0) (1,-2) (3,0) (3,-2)}, draw, dashed] (native) {};
    
    \draw (-1, -2.3) node[]{Java};
    \draw (2, -2.3) node[]{Native};
\end{tikzpicture}
        }
        \caption{Getter}
        \label{fig:four_cases:getter}
    \end{subfigure}
    \begin{subfigure}{.49\linewidth}
        \centering
        \resizebox{.9\linewidth}{!}{
            \begin{tikzpicture}
    \tikzstyle{arrowStyle}=[-latex]
    \tikzset{node/.style={minimum height=0.9cm, rounded corners, text width=1.9cm,align=center,scale=.8}}

    \path
    (-1,-.5) node[node,draw] (source) {  = source()}
    ++(0,-1) node[node,draw] (native) {native( )}
    ++(3,.5) node[node,draw] (sink) {sink( )}
    ;
    
    \node[circle,draw=black, fill=black, inner sep=0pt,minimum size=3pt] (a) at (-1,.5) {};
    
    \node[circle,draw=red, fill=red, inner sep=0pt,minimum size=3pt] (a) at (2.22,-1) {};
    \node[circle,draw=red, fill=red, inner sep=0pt,minimum size=3pt] (b) at (-1.65,-.5) {};
    \node[circle,draw=red, fill=red, inner sep=0pt,minimum size=3pt] (d) at (-.68,-1.5) {};
    
    \draw[->,>=latex,red] (b) to[out=290,in=90] (d);
    \draw[->,>=latex,red] (d) to[out=270,in=120] (a);
    
    \draw[->,>=latex,black] (-1,.5) -- (source);
    
    \draw[->,>=latex,black] (source) -- (native);
    
    \draw[->,>=latex,densely dotted] (native.north east) to[out=20,in=100] (2,.1);
    
    \node[fit={(-2,0) (-2,-2) (0,0) (0,-2)}, draw, dashed] (java) {};
    \node[fit={(1,0) (1,-2) (3,0) (3,-2)}, draw, dashed] (native) {};
    
    \draw (-1, -2.3) node[]{Java};
    \draw (2, -2.3) node[]{Native};
\end{tikzpicture}
        }
        \caption{Leaker}
        \label{fig:four_cases:leaker}
    \end{subfigure}
    \begin{subfigure}{.49\linewidth}
        \centering
        \resizebox{.9\linewidth}{!}{
            \begin{tikzpicture}
    \tikzstyle{arrowStyle}=[-latex]
    \tikzset{node/.style={minimum height=0.7cm, rounded corners, text width=1.9cm,align=center,scale=.8}}

    \path
    (-1,-.3) node[node,draw] (source) {  = source()}
    ++(0,-.7) node[node,draw] (native) { = native( )}
    ++(0,-.7) node[node,draw] (sink) {sink( )}
    ++(3,.7) node[node,draw] (return) {return}
    ;
    
    \node[circle,draw=black, fill=black, inner sep=0pt,minimum size=3pt] (a) at (-1,.5) {};
    
    \node[circle,draw=red, fill=red, inner sep=0pt,minimum size=3pt] (a) at (-1.65,-1) {};
    \node[circle,draw=red, fill=red, inner sep=0pt,minimum size=3pt] (b) at (-.55,-1) {};
    \node[circle,draw=red, fill=red, inner sep=0pt,minimum size=3pt] (c) at (-1.65,-.3) {};
    \node[circle,draw=red, fill=red, inner sep=0pt,minimum size=3pt] (d) at (2.5,-1) {};
    \node[circle,draw=red, fill=red, inner sep=0pt,minimum size=3pt] (e) at (-.78,-1.7) {};
    
    \draw[->,>=latex,red] (c) to[out=350,in=110] (b);
    \draw[->,>=latex,red] (b) to[out=20,in=90] (d);
    \draw[->,>=latex,red] (d) to[out=270,in=340] (a);
    \draw[->,>=latex,red] (a) to[out=270,in=90] (e);
    
    \draw[->,>=latex,black] (-1,.5) -- (source);
    
    \draw[->,>=latex,black] (source) -- (native);
    \draw[->,>=latex,black] (native) -- (sink);
    
    \draw[->,>=latex,densely dotted] (native.north east) to[out=20,in=90] (2,.1);
    
    \node[fit={(-2,0) (-2,-2) (0,0) (0,-2)}, draw, dashed] (java) {};
    \node[fit={(1,0) (1,-2) (3,0) (3,-2)}, draw, dashed] (native) {};
    
    \draw (-1, -2.3) node[]{Java};
    \draw (2, -2.3) node[]{Native};
\end{tikzpicture}
        }
        \caption{Proxy}
        \label{fig:four_cases:proxy}
    \end{subfigure}
    \begin{subfigure}{.49\linewidth}
        \centering
        \resizebox{.9\linewidth}{!}{
            \begin{tikzpicture}
    \tikzstyle{arrowStyle}=[-latex]
    \tikzset{node/.style={minimum height=0.9cm, rounded corners, text width=1.9cm,align=center,scale=.8}}

    \path
    (-1,-1) node[node,draw] (native) {native()}
    ++(3,.5) node[node,draw] (source) { = source()}
    ++(0,-1) node[node,draw] (sink) {sink( )}
    ;
    
    \node[circle,draw=black, fill=black, inner sep=0pt,minimum size=3pt] (a) at (-1,.5) {};
    
    \node[circle,draw=red, fill=red, inner sep=0pt,minimum size=3pt] (a) at (1.3,-.5) {};
    \node[circle,draw=red, fill=red, inner sep=0pt,minimum size=3pt] (c) at (2.23,-1.5) {};
    
    \draw[->,>=latex,red] (a) to[out=270,in=90] (c);
    
    \draw[->,>=latex,black] (-1,.5) -- (native);
    
    \draw[->,>=latex,black] (source) -- (sink);
    
    \draw[->,>=latex,densely dotted] (native.north east) to[out=20,in=90] (2,.1);
    
    \node[fit={(-2,0) (-2,-2) (0,0) (0,-2)}, draw, dashed] (java) {};
    \node[fit={(1,0) (1,-2) (3,0) (3,-2)}, draw, dashed] (native) {};
    
    \draw (-1, -2.3) node[]{Java};
    \draw (2, -2.3) node[]{Native};
\end{tikzpicture}
        }
        \caption{Delegation}
        \label{fig:four_cases:delegation}
    \end{subfigure}
    \caption{Four propagation scenarios through native code}
    \label{fig:four_cases}
\end{figure}
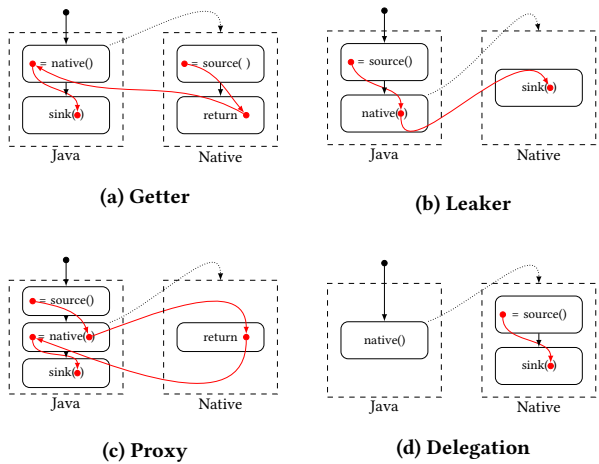

Note that "Source/Sink in native code" means that the call to a sensitive method is actually performed in native code, but the sensitive method is always a method from the Android framework accessed with JNI (e.g., calling with JNI the \texttt{getDeviceId()} from the native code).
For each of these cases, at least one step happens in native.
Figure~\ref{fig:four_cases} illustrates these four cases.
The red dots represent tainted information from a source method, and the red arrows represent how this information flows in the program.
The \emph{Getter} use-case allows developers to get sensitive data from the native code to leak it in the Java world.
The \emph{Leaker} use-case allows developers to get sensitive data from the Java world to leak it in the native world.
Regarding the \emph{Proxy} use-case, the sensitive information is retrieved in the Java world, sent to the native world to "break" the flow, and sent back to the Java world to be leaked.
Concerning the \emph{Delegation} use-case, a simple native function is called from the Java world, and the sensitive information is retrieved and leaked in the native world.

Our benchmark apps has been built, upon these four cases that we identified, to be representative of these cases, with combination of multiple cases.

\textbf{Results}: Table~\ref{table:bench_results} provides the results of our experiments.
\flowdroid is clearly limited and not designed to handle native code. Therefore its inferior performances are not surprising.
Indeed, \flowdroid gets a precision and recall of 0\% on this benchmark.

\begin{table}[h]
{\begin{center}\small$\ostar$ = true-positive, \textcolor{red}{$\star$} = false-positive,         \textcolor{red}{$\bigcirc$} = false-negative\end{center}}
    \begin{adjustbox}{width=\columnwidth,center}
        \begin{tabular}{lccc}
        \hline
        \rowcolor{gray!70}
        \bf Test Case & \bf Leak & \bf \flowdroid  & \bf \jucify \\
        \hline \hline
        \rowcolor{gray!30}
        getter\_imei & $\bullet$ & \textcolor{red}{$\bigcirc$}  &  $\ostar$ \\
        leaker\_imei & $\bullet$ &  \textcolor{red}{$\bigcirc$} & $\ostar$  \\
        \rowcolor{gray!30}
        proxy\_imei & $\bullet$ &  \textcolor{red}{$\bigcirc$} & $\ostar$  \\
        delegation\_imei & $\bullet$ &  \textcolor{red}{$\bigcirc$} &  $\ostar$ \\
        \rowcolor{gray!30}
        getter\_string & $\circ$ &   & \textcolor{red}{$\star$}  \\
        leaker\_string & $\circ$ &   & \textcolor{red}{$\star$}  \\
        \rowcolor{gray!30}
        proxy\_double & $\bullet$ &  \textcolor{red}{$\bigcirc$} &  $\ostar$ \\
        delegation\_proxy & $\bullet$ & \textcolor{red}{$\bigcirc$}  &  $\ostar$ \\
        \rowcolor{gray!30}
        getter\_leaker & $\bullet$ & \textcolor{red}{$\bigcirc$}  & $\ostar$  \\
        getter\_proxy\_leaker & $\bullet$ & \textcolor{red}{$\bigcirc$}  & $\ostar$  \\
        \rowcolor{gray!30}
        getter\_imei\_deep & $\bullet$ & \textcolor{red}{$\bigcirc$}  & $\ostar$  \\
        \rowcolor{gray!50}
        \multicolumn{4}{c}{Sum, Precision, Recall} \\
        $\ostar$, higher is better & & 0 & 9 \\
        \rowcolor{gray!30}
        \textcolor{red}{$\star$}, lower is better & & 0 & 2 \\
        \textcolor{red}{$\bigcirc$}, lower is better & & 9 & 0 \\
        \rowcolor{gray!30}
        Precision $p = \ostar / (\ostar + $ \textcolor{red}{$\star$} $)$ & & 0\% & 82\% \\
        Recall $r = \ostar / ( \ostar + $ \textcolor{red}{$\bigcirc$} $)$ & & 0\% & 100\% \\
        \rowcolor{gray!30}
        $F_1$-score $= 2pr/(p+r)$ & & 0\% & 90\% \\
        \hline
        \end{tabular}
    \end{adjustbox}
    \caption{Results of data leak detection through native code in bench apps. \flowdroid column represents the results of running \flowdroid alone. \jucify column represents the results of running \flowdroid after applying \jucify}
    \label{table:bench_results}
\end{table}

Nevertheless, we can see that after applying \jucify, \flowdroid performance is significantly boosted.
Indeed, it can detect all the leaks present in the benchmark, hence achieving a recall score of 100\%.
Regarding apps \emph{getter\_string} and \emph{leaker\_string}, \flowdroid reports for both of them a false positive alarm leading to a precision of 82\% on this benchmark.
In these apps, a string is sent outside the apps, not sensitive data.
This is easily explained by the fact that when \jucify reconstructs the native function's behavior, 
it uses opaque predicates to approximate what variable can be returned by the current function given its signature.
Therefore, there is a path in which the sensitive data is considered, whereas it is not leaked.

\highlight{
\textbf{RQ3 answer:}
Jucify is essential for boosting state-of-the-art static analyzers such as FlowDroid to take into account native code. On our constructed benchmark, FlowDroid, which failed to discover any leak, is now able to precisely identify leaks in a high number of samples (F1-score at 90\%).
}

\subsection{RQ4: \jucify in the wild}
\label{sec:eval:rq4}

In this section, we evaluate \jucify in the wild from two points of view: \ding{182} a quantitative assessment in section~\ref{sec:evaluation:call_graph_augmentation_wild}; and \ding{183} a qualitative assessment in section~\ref{sec:evaluation:leaks_real_world}.

\subsubsection{RQ4.a: To what extent can \jucify augment apps' call-graphs and reveal previously unreachable Java methods?}\hfill
\label{sec:evaluation:call_graph_augmentation_wild}

To assess to what extent call-graphs are augmented by \jucify, we applied it on two sets of Android apps: 1) \num{1000} benign apps; 2) \num{1000} malware. 
Note that we only selected apps that contain at least one \texttt{.so} file.
The results reported concern apps for which \jucify succeeded to make call-graph changes.
The reasons for which there are apps without changes is related to the absence of bytecode-to-native links (i.e., for 559 goodware and 384 malware) and/or \jucify reaching the 1h-timeout (i.e., for 15 goodware and 51 malware).

\textbf{Number of nodes and edges in call-graphs:} 
We first report the average number of nodes (i.e., the number of methods) and edges (i.e., the number of potential invocations) in the call-graphs obtained before and after having applied \jucify.

The call-graph augmentations brought by \jucify are visible in Table~\ref{table:callgraph}.
Column \emph{\# apps} represents the number of apps for which \jucify made callgraph changes, i.e., they did not reach the timeout and contained bytecode-native links.
We notice that about half of the apps' call-graphs are impacted by \jucify (426 and 565 for goodware and malware respectively).
We then notice that the number of nodes and edges added by \jucify is higher for goodware than for malware: 270 vs. 197 on average per app for nodes, and 778 vs. 446 for edges.
This shows that classical static analyzers that do not take into account the native code, overlook a significant amount of nodes and edges in their call-graph.

\begin{table}[ht!]
      \begin{adjustbox}{width=\columnwidth,center} \begin{tabular}{lc||c|c||c|c||c|c}
         &  & \multicolumn{2}{c||}{\textbf{Before \jucify}} & \multicolumn{2}{c||}{\textbf{After \jucify}} &
           \multicolumn{2}{c}{\textbf{Difference}} \\
            \cline{3-8}
            & \multicolumn{1}{c||}{\# apps} & \# Nodes & \# Edges &  \# Nodes  & \# Edges &  Added Nodes  & Added Edges\\
            \hline
            Goodware & \multicolumn{1}{|c||}{426} & \num{4515} & \num{18287} & \num{4784} & \num{19065} & \num{270} (+5.9\%) & \num{778} (+4.2\%) \\ \hline
            Malware & \multicolumn{1}{|c||}{565} & \num{3056} & \num{14266} & \num{3253} & \num{14712} & \num{197} (+6.4\%) & \num{446} (+3.1\%) \\ 
        \end{tabular}
        \end{adjustbox}
    \caption{Average numbers of nodes and edges before and after \jucify on 426 goodware and 565 malware}
    \label{table:callgraph}
\end{table}

\begin{figure}[ht!]
    \centering
    \includegraphics[scale=.5]{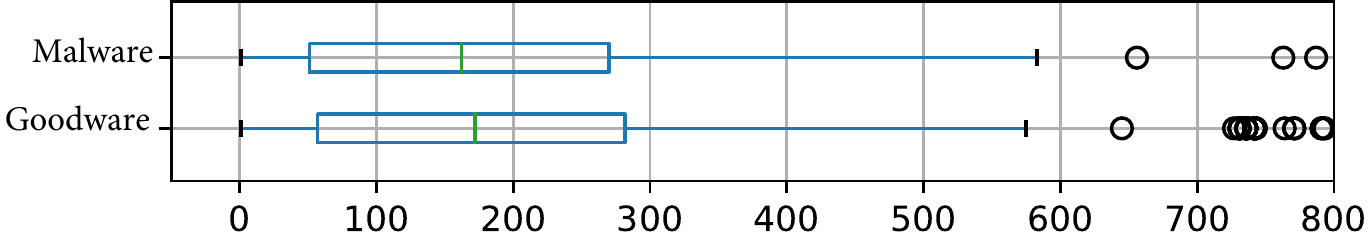}
    \caption{Distribution of the number of binary functions nodes in benign and malicious Android apps}
    \label{fig:new_binary_nodes}
\end{figure}

\textbf{Number of binary functions in the augmented call-graph:}
Newly added nodes can be explained by the binary functions (i.e., functions in the native code part) that are now considered in the unified call graph yielded by \jucify.
Figure~\ref{fig:new_binary_nodes} details the distributions of the number of binary functions for both datasets.
We notice that benign apps tend to have more added binary function nodes (median = 172, and mean = \num{269.7}) in the call-graph than malicious apps (median = 162, and mean = \num{197.2}).
Both distributions are significantly different, as confirmed by a Mann-Whitney-Wilcoxon (MWW) test~\cite{mann1947} (significance level set at 0.05).

\begin{figure}[ht!]
    \centering
    \includegraphics[scale=.5]{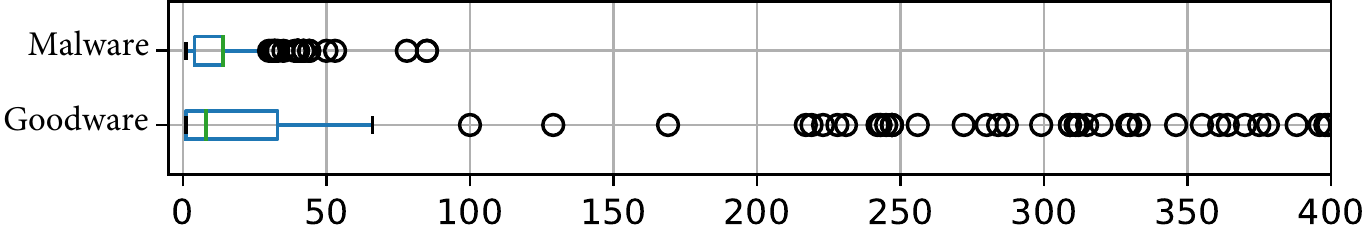}
    \caption{Distribution of the number of bytecode-to-native edges in benign and malicious Android apps}
    \label{fig:java_to_native}
\end{figure}

\textbf{Number of bytecode-to-native call-graph edges:}
Newly created edges can originate from native function invocations in bytecode methods  (i.e., entry invocations). 
We compute the number of bytecode-to-native edges in apps' call-graph and detail their distributions over our datasets in Figure~\ref{fig:java_to_native}.
The difference between malware and goodware is significant, with a median equal to 14 for malware and 8 for goodware.
Overall, \jucify reveals a total of \num{6758} bytecode-to-native invocations in the malware dataset and \num{29908} in the goodware dataset.

\begin{figure}[ht!]
    \centering
    \includegraphics[scale=.5]{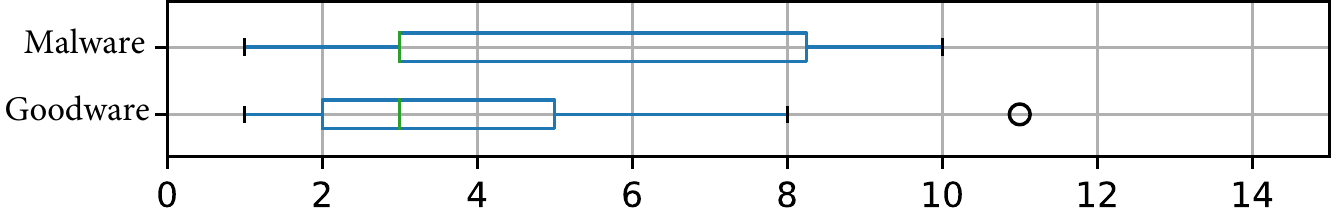}
    \caption{Distribution of the number of native-to-bytecode edges in benign and malicious Android apps}
    \label{fig:native_to_java}
\end{figure}

\textbf{Number of native-to-bytecode call-graph edges:}
Newly add\-ed edges can also originate from bytecode methods invoked in native functions (i.e., exit invocations with reflection-like mechanisms as explained in Section~\ref{sec:background:jni:nativetojava}).
The median of number of edges is significantly low for both goodware and malware.
Indeed, the median of native-to-bytecode edges is equal to 3 for both datasets, the distribution is available in Figure~\ref{fig:native_to_java}.
Overall, \jucify reveals a total of $261$ native-to-bytecode invocations in the entire goodware set and $4288$ in the malware set.
The conclusion that can be drawn from these results is the following:

the low numbers of native-to-bytecode edges in goodware shows that benign apps make little use of reflection-like mechanisms to invoke Java methods from native code, compared to malware.

\textbf{New previously unreachable bytecode methods:}
By considering native code, \jucify can reveal previously unreachable bytecode methods that are now reachable (because called from the native part). 
The number of previously unreachable bytecode methods is highly linked to the number of native-to-bytecode call-graph edges discussed in the previous paragraph. 
However, a new edge from native to bytecode can simply end to a previously reachable node, which does not present an interest here.
Indeed, newly reachable nodes are interesting since they allow static analyzers to not consider them as dead code anymore.
In Section~\ref{sec:evaluation:data_flow}, we give a concrete example of the importance of this metric.

Overall, \jucify can reveal \num{34} previously unreachable bytecode methods in 18 benign apps (with a maximum of 5 for one given app).
For malicious apps, \jucify reveals \num{122} previously unreachable bytecode methods called from native code in 54 apps.
This accounts for 13\% of native-to-bytecode invocation in goodware and 2.8\% for malware.
This suggests that in most cases when Android app developers invoke bytecode methods from native code, it is to trigger bytecode methods that are already reachable from the bytecode.
However, this shows that a non-negligeable proportion of bytecode invocation from the native in goodware and malware are overlooked by classical static analyzers since they account for non-reachable nodes in original bytecode callgraph.

\textbf{Goodware vs. Malware native/bytecode calls:}
To better understand the difference between goodware and malware, we inspected the native functions invoked from the bytecode and the bytecode methods invoked from the native code.
Results indicate that in \num{82.7}\% of the cases, the native function \texttt{Java\_mono\_android\_\-Runtime\_register} is invoked from the bytecode in goodware.
In fact, most of the top invoked native functions in goodware are from the \texttt{mono} framework, which is used by \textsc{Xamarin}~\cite{xamarin}.
The same method is, however, not found in the malware dataset.
The top invoked native functions in malware is composed of different elements such as \texttt{Java\_\-com\_\-seleuco\_\-mame4all\_\-Emulator\_\-setPadData}, \texttt{Java\_\-com\_\-shunpay210\_\-sdk\_\-CppAdapter210\_pay}, or more suspicious functions: \texttt{Java\_\-iqqxF\_\-TZfff\_\-ggior} and \texttt{Java\_\-glrrx\_\-efgnp\_\-twCJN}.

From native to bytecode, we note some interesting insights:
while benign apps invoke from the native code, in the majority of cases, bytecode methods like \texttt{Context.get\-Package\-Name} (\num{14.2}\%), or \texttt{Thread\-Local.get} (\num{8.2}\%), malicious apps invoke methods such as \texttt{Telephony\-Manager.get\-Device\-Id} (\num{2.4}\%), or \texttt{Telephony\-Manager.\-getSubscriberId} (\num{4.3}\%) which can indicate suspicious behaviors.

Our results become more convincing by focusing on bytecode methods that were previously unreachable in call-graphs and called from native code. 
While most of the bytecode methods that were previously unreachable and called in the native code in goodware are \texttt{Mono} framework methods, in malware, the situation is different.
Indeed, the most used bytecode methods in native code are dedicated to payment libraries (e.g., \texttt{com.\-shunpay208.sdk.\-Shun\-Pay\-208}), and sensitive methods such as \texttt{getDeviceId}.

\highlight{
\textbf{RQ4.a answer:}
\jucify helps to discover new paths in app behaviour.
It augments call-graphs with about 5-6\% new nodes in both benign and malware apps. Overall, apps tend to use much more bytecode-to-native invocations than native-to-bytecode.
However, malware seems to use bytecode invocations from native to perform suspicious activities.
}

\subsubsection{RQ4.b: Can \jucify reveal previously unreachable sensitive data leaks that pass through native code in real-world apps?}\hfill\\
\label{sec:evaluation:leaks_real_world}

With this RQ, our goal is to assess \jucify from a qualitative point of view. In particular, we check whether the call-graphs augmented by \jucify with previously unseen nodes are relevant. To that end, we run \jucify and \flowdroid on  real-world apps to check if \flowdroid can detect sensitive data leaks through the native code.

\noindent
\textbf{Experimental setup:}
To assess \jucify in the wild, we selected malicious applications since the intuition is that malicious apps tend to leak sensitive data more than goodware.
Therefore, we randomly selected \num{1800} malicious apps (i.e., VirusTotal score $>$ 20) from Androzoo~\cite{Allix:2016:ACM:2901739.2903508} that contain .so files.
Besides, to detect data leaks, we used the default sources and sinks provided by \flowdroid.
For each of these \num{1800} apps, we set a 1-hour timeout (30 min for the symbolic execution and 30 min for \flowdroid).

\noindent
\textbf{Findings:}
Among the \num{1800} malicious apps, \num{1460} contained Java native methods declaration(s) in the code.
In total, \jucify was able to augment the call-graph of \num{1066} (i.e., 73\%) of the \num{1460} apps that contain both .so files and Java native method declaration in bytecode.
From these \num{1460} apps, \flowdroid revealed sensitive data leaks that take advantage of the native code in 14 apps.
These 14 apps were manually checked and confirmed to contain sensitive data leaks that goes through the native code.
Note that this number is highly linked to the source and sink methods used.

In the following, we discuss two case studies where \jucify was able to reveal sensitive data leaks that pass through native code. Both Android apps were manually checked by the authors to confirm the presence of a leak detected by \flowdroid.

\subsubsection{Getter-Scenario Case Study}
\label{sec:eval:getter}

In Figure~\ref{fig:four_cases:getter} we illustrated an example of how malware developers can rely on native code to hide, from static analyzers, the \underline{\em retrieval} of sensitive data from static analyzers.
\jucify revealed an Android malware~\footnote{\scriptsize SHA-256: 54DAFDF3635B18C0FD9F5CE89FE14C072D75AB4687B376FBADF370388574DC14} implementing this specific behavior.
\jucify reconstructed the \texttt{A()} native method from the \texttt{com.y} class as the following: "\texttt{<DummyBinaryClass: java\-.lang\-.String Java\_com\_y\_A(android.content.Context)>}".
In this native function, the IMEI number of the device is obtained via the JNI interface and returned as a result.
This reconstructed method is called in method \texttt{b()} of class \texttt{com.cance.b.q} to store the IMEI number.
The resulting IMEI number is then wrapped and transferred to a method to log it.

After examining the VirusTotal report of this app, we found that the flags raised by antiviruses refer to Trojan behavior and explicitly mention the retrieval of sensitive information from the device as well as the use of native code in the implementation of the malicious behaviour. To some extent, this corroborates that \jucify contributed to uncover a malicious behaviour that is hidden through exploiting native-to-bytecode links (which state of the art static analyzers could not be aware of).

\subsubsection{Leaker-Scenario Case Study}

In Figure~\ref{fig:four_cases:leaker}, we illustrated how app developers can rely on native code to hide the \underline{\em leakage} of sensitive data.
\jucify revealed an Android malware~\footnote{\scriptsize SHA-256: A0B7BFBC272B462A2F59CC09ACC8B75114137CF7A2B391201C14C1A90EA7E369} with this behavior.

First, the IMEI number is obtained in the \texttt{getOperator()} method of the \texttt{com.umeng.adutils.AppConnect} class and stored in the \texttt{imei} field of the same class.
Then, in the \texttt{processReplyMsg()} method of this class (method which is
triggered when an SMS is received), the IMEI number is wrapped in another string and sent to the native method "\texttt{stringFromJNI()}" as a parameter.
\jucify's instrumentation engine constructed the following method from this native method: "\texttt{<DummyBinaryClass: java.lang.String Java\_\-com\_\-umeng\_adutils\_SdkUtils\_stringFromJNI(\-and\-roid\-.\-app\-.\-P\-endingIntent,java.lang.String,java.lang.String)>}".
This latter has been populated with the information given by the symbolic execution and revealed that the \texttt{sendText\-Message()} met\-hod from the \texttt{android.telephony.Sms\-Manager} cl\-ass is called with the valued derived from the IMEI number as parameter.

To summarize, a value derived from the IMEI number is sent out of the device using an SMS through the native code.
Doing so, the leak would have remained undetected without \jucify.

As in the previous case study, we examined the VirusTotal report of this app. In their majority, antiviruses flag it as a Trojan app. Some reports even explicit tag the use of \texttt{getDeviceId()} and of native code for the malicious operations. Thus, with \jucify we enabled an existing analyzer to uncover a leak being performed through native code.

\highlight{
\textbf{RQ4.b answer:}
\jucify is effective for highlighting data flows across native code that were previously unseen.
Indeed, its enhanced call-graphs enable static analysers to reveal sensitive data leaks within real-world Android apps.
}
\section{Limitations}
\label{sec:limitations}
Our approach is a step towards realizing the ambition of full code unification for Android static analysis.
Our current prototype of \jucify, despite promising performances, presents a few limitations: First, our implementation relies on existing tools to extract native call-graphs and mutual invocations between bytecode and native code. Limitations of these tools are therefore carried over to \jucify.
Such limitations include the exponential analysis time for symbolic execution, the limitation in finding the boundaries of native functions, the unsoundness in app modeling with \flowdroid due to reflective calls~\cite{li2016droidra}, multi-threading~\cite{maiya2014race}, and dynamic loading~\cite{xue2017auditing}.

Second, our prototype currently relies on symbolic execution which is known to be non-scalable in the general case.
Therefore, as described in Section~\ref{sec:evaluation:call_graph_augmentation_wild}, the call-graph of some Android apps was not augmented due to the symbolic execution that did not return native-bytecode links and/or due to the timeout.

Third, a major limitation of \jucify lies in the fact that it does not yet reconstruct native functions behavior with high precision. 
Indeed, as described in Section~\ref{sec:cg_to_jimple}, for the native functions that represent Java native function,
\jucify considers a partial list of statements:  
it employs opaque predicates to guide static analyzers into considering every possible path during analyses.
Moreover, \jucify overlooks native functions that are not explicitly targeted by JNI Java calls since it cannot approximate their behavior in the current implementation.
As a result, \jucify cannot generate native functions' control flow graphs with Jimple statements that cover the full behavior of functions. 
This limitation implies that if, for instance, a leak is performed by using Internet communication implemented "purely" in C (e.g., with a socket), then this leak would not be detected with \flowdroid even after \jucify processing.
Also, during the reconstruction phase described in Section~\ref{sec:cg_to_jimple}, in some cases where the number of parameter is important, the number of parameter combination can explode.
This can lead to methods being extremely long that might not represent reality.
We plan to address this limitation in future work.
\section{Threats to validity}
\label{sec:threat_to_validity}

\textbf{Manual Checking.}
To check the correctness of the results, we manually checked a hundred 
Android apps.
To do so, we relied on Java bytecode decompilers and native code decompilers such as Jadx~\cite{jadx} and RetDec~\cite{retdec}.
Although native code manual checking is challenging, we were able to confirm that the native nodes added by \jucify matched the nodes from the native callgraph constructed by \nativediscloser.
Regarding bytecode-to-native links, as the symbols were always available for the apps we checked (since native methods were statically registered), we were able to confirm the correctness of those links in the callgraph generated by \jucify.
We reverse-engineered these apps and were able to reach the same conclusions.
Regarding native-to-bytecode links, the method names are represented as strings, which are not directly available in the native code.
Therefore, we faced a challenge to check if the symbolic execution yielded correct links.
One way to verify would be to execute the code part to trigger the native code and ensure that the correct information are yielded by \nativediscloser, but this is a challenge per se and it is out of the scope of this study.
Therefore, we made the hypothesis that the symbolic yields correct results.
\section{Related work}
\label{sec:related_work}

\noindent
{\bf Static analysis of Android apps.}
Static analysis of Android apps is widely explored to assess app properties.
Less than 10 years after the introduction of Android, a systematic literature review~\cite{LI201767} has shown that over one hundred papers presented static approaches to analyze Android apps. The review highlights that Android apps' security vetting is one of the main concerns for analysts, who assess properties such as sensitive data leak detection~\cite{arzt2014flowdroid, li2015iccta, samhi2020raicc}, or check for maliciousness~\cite{li2015potential,grace2012riskranker,wu2012droidmat}.
Static approaches have also been implemented to identify functional and non-functional defects~\cite{wu2016static, cruz2019energy} and towards fixing runtime crashes~\cite{kong2019mining,tan2018repairing}. Static analysis is also further leveraged to collect information in apps towards improving dynamic testing approaches~\cite{kong2018automated, mao2016sapienz,zhang2016automated,su2017guided}. Given these fundamental usages of static analysis, it is essential to take into account all code that implements any part of the app behavior. Therefore, the fact that many analyses are reduced to focus on the bytecode (while leaving out native code within app packages) constitutes a severe threat to validity in many studies.

\noindent
{\bf Binary analysis.}
Binary analysis techniques have been applied for different platforms, using static~\cite{10.1145/1127577.1127590,805197,5726949,feist2014statically}, dynamic~
\cite{bayer2006dynamic,li2013dynamic,choi2015dynamic}, hybrid~\cite{roundy2010hybrid,damodaran2017comparison,8530021} and machine-learning-based~\cite{lee2017learning,8094438,maier2019typeminer,7582748} approaches.
A recent work~\cite{george2020native} tackles the challenging task of analyzing binaries by combining declarative static analysis (using Datalog declarative logic-based programming language) with reverse-engineering techniques to perform x-refs analysis in native libraries using Radare2~\cite{radare2}.
In the Android realm, analysis of binaries can be essential to cope with obfuscation~\cite{kan2019automated}.

\noindent
{\bf Cross-language analysis.}
Several researchers have also acknowledged the presence of native code alongside bytecode in their analysis of Android apps.
For instance, in 2016, Alam et al.~\cite{ALAM2017230} presented \emph{DroidNative} which can perform Android malware detection considering both the bytecode and the native code.
\emph{NDroid}~\cite{6903578} and \emph{TaintArt}~\cite{10.1145/2976749.2978343} were proposed for dynamic taint analysis to track sensitive information flowing through JNI.
 \emph{JN-SAF}~\cite{10.1145/3243734.3243835} is also proposed as an inter-language static analysis framework to detect sensitive data leaks in Android apps, taking into account native code.
 All the aforementioned tools, however, are task-specific. They also, typically, perform their analyses separately for bytecode and native code, and later post-process and merge the outputs to present unified analysis results. In contrast, \jucify proposes to unify the representation before task-specific analyses. This enables other analyses to be built upon the output of \jucify.
For experimental assessment of \jucify representation for data flow analysis (RQ-5), we envisioned a comparison with JN-SAF. Unfortunately, two co-authors independently failed to run the tool. 

Overall, there are various approaches and studies~\cite{9286029,RizzoPhd,afonso2016going,10.1145/2627393.2627396} in the literature that investigate the possibility to analyze apps by account for the different language-specific artifacts in the package. 
Although the approaches described are promising for cross-language analysis, they do not generally offer a practical framework to unify the representation of both the bytecode and the native code into a single model that standard static analysis pipelines can leverage.
Our prototype \jucify does bring such a unified model and targets the Jimple intermediate representation, which is the default internal representation of \soot. Therefore, by pushing in this research direction, we expect to provide the community with a readily usable framework, which will allow to (re)perform their analyses on whole code in Android apps.

\section{Conclusion}
\label{sec:conclusion}

We contribute in the ambitious research agenda of unifying bytecode and native code to support comprehensive static analysis of Android apps. We presented \jucify, as a significant step towards this unification: it generates a native call-graph that is merged with the bytecode call-graph based on links retrieved via symbolic execution. 
In this model (i.e., the unified call-graph), we are able to heuristically populate specific native functions with Jimple statements.
The Jimple intermediate representation was selected to readily support existing static analysers based on the Soot framework.

We first empirically showed that \jucify significantly improves Android apps call-graphs, which are augmented (to include native code nodes) and enhanced (to reveal previously unreachable methods). 
Then, we showed that \jucify holds its promise in supporting state-of-the-art analyzers such as \flowdroid in enhancing their taint tracking analysis.
Finally, we discuss how \jucify can reveal sensitive data leaks that pass through the native code in real-world Android apps, which were previously undetectable.
\section{Data Availability}
\label{sec:data_availability}
For the sake of Open Science, we provide to the community all the artifacts used in our study.
In particular, we make available the datasets used for our experimentations, the source code of \jucify, the \jucify executable, and our benchmark apps.
All artifacts (code, benchmarks, results) are available at:%
\begin{center}
    \url{https://github.com/JordanSamhi/JuCify}
\end{center}
\section{Acknowledgment}
\label{sec:acknowledgment}
This work was partly supported 
(1) by the Luxembourg National Research Fund (FNR), under projects Reprocess C21/IS/16344458 the AFR grant 14596679,
(2) by the SPARTA project, which has received funding from the European Union's Horizon 2020 research and innovation program under grant agreement No 830892, 
(3) by the NATURAL project, which has received funding from the European Research Council under the European Union’s Horizon 2020 research and innovation programme (grant N° 949014),
and (4) by the INTER Mobility project Sleepless@Seattle No 13999722.

\bibliographystyle{ACM-Reference-Format}

\bibliography{bib}

\end{document}